\documentstyle[preprint,aps]{revtex} 
\begin{document}
\draft
\date{to appear in {\sl The Philosophical Magazine B}}
\title{Optical absorption of non-interacting tight-binding electrons
in a Peierls-distorted chain at half band-filling}
\author{F.~Gebhard\footnote{e-mail: {\tt florian\verb2@2gaston.ill.fr}}}
\address{ILL Grenoble, B.~P.\ 156x, F-38042 Grenoble Cedex 9, France}
\author{K.~Bott, M.~Scheidler, P.~Thomas, S.~W.~Koch}
\address{Dept.~of Physics and Materials Sciences Center,\\
Philipps University Marburg, D-35032~Marburg, Germany}

\maketitle

\begin{abstract}%
In this first of three articles
on the optical absorption of electrons in
half-filled Peierls-distorted chains we present 
analytical results for non-interacting tight-binding electrons.
We carefully derive explicit expressions for the current operator,
the dipole transition matrix elements, and the optical absorption
for electrons with a cosine dispersion relation of band width~$W$
and dimerization parameter~$\delta$.
New correction (``$\eta$''-)terms to the current operator
are identified.
A broad band-to-band transition is found in the frequency range
$W\delta < \omega < W$ whose shape is determined
by the joint density of states for the upper and lower Peierls subbands
and the strong momentum dependence of the transition matrix elements.

\vskip2cm\noindent
PACS1996: 71.20.Rv, 71.10.Ca, 36.20.Kd
\end{abstract}

\newpage
\section{Introduction}

Polymers and charge-transfer salts are examples
for almost ideal one-dimensional materials (Farges 1994).  %cite
Besides their potential technological applications
these systems play an important test ground for
our theoretical understanding of electron-electron and
electron-phonon interactions in low dimensions.
The theoretical study of the corresponding one-dimensional
models instead of their three-dimensional counterparts
has the further advantage that much larger systems
can be handled numerically, and some of the conceptually simplest can even
be solved analytically.

If one wants to show that a model is indeed appropriate
for a real material one has to calculate experimentally accessible quantities.
Analytical calculations often require the introduction of further
(uncontrolled) approximations while the analysis of data from numerical
approaches is hampered by finite size effects. 
This makes the assessment of the model quality a difficult task.
Even if a model can be solved 
without resorting to uncontrolled approximations
the interpretation and relevance
of the results for real materials remains a matter of debate.
Theoretical investigations concentrate on
ideal systems, e.g., infinitely extended and perfectly homogeneous, 
that cannot be made in practice.
Thus it should be kept in mind that experimental results may
dominantly show effects of disorder and interactions with
the chain environment which are usually neglected
in theoretical approaches.

Despite these conceptual problems the 
analysis of basic models remains the first task
to be accomplished in theory. In principle, models
will differ in their experimental predictions
which will allow to discriminate between them.
In addition, model parameters are usually fitted
by comparison with experiment.

It is widely believed that polyacetylene and other polymers
are basically described
by independent electrons coupled to a linear chain of atoms
(Heeger, Kievelson, Schrieffer and Su 1988);
(Baeriswyl, Campbell and Mazumdar 1992); (Schott and Nechtschein 1994).  %cite
For a fixed lattice distortion the generic model is
the dimerized  tight-binding (or H\"{u}ckel) model.
The optical absorption for this independent
free electron model was already addressed in the past 
(Baeriswyl, Harbeke, Kiess, Meier, and Meyer 1983); (Abe 1993).    %cite
In this work we comprehensively derive analytical
expressions for the dipole matrix elements and the optical absorption
in the tight-binding approximation.
In contrast to previous works we take
special care of the derivation of the current and
the dipole operator. 

Some charge-transfer salts are understood as strongly correlated
one-dimensional electron systems for half-filled bands
(Mott-Hubbard insulators)
(Alc\'{a}cer, Brau, and Farges 1994).   %cite
The extended Hubbard model for interacting electrons
on a distorted chain at half-filling is considered appropriate for these
materials (Mazumdar and Dixit 1986); (Fritsch and Ducasse 1991); 
(Mila 1995).   %cite
The theoretical study of these model systems is subject to
two forthcoming articles
(Gebhard, Bott, Scheidler, Thomas, and Koch 1996).  %cite

The paper is organized as follows.
In section~\ref{Hamiltpei} we set up the Hamiltonian and
diagonalize it. In section~\ref{currentop} we determine the current operator
for tight-binding electrons on a Peierls-distorted chain.
We show that the distortion gives rise to
correction terms that have been neglected in the literature.
In section~\ref{appeleccurr} we derive the standard formulae
for the optical conductivity.
In section~\ref{peierlssec} we present 
explicit expressions for the optical absorption
of free, independent tight-binding electrons 
moving on a Peierls-dimerized chain at half band-filling.
In section~\ref{appdipole} we show how to derive 
the results via perturbation theory
using the regular part of the electrical dipole operator.
This approach is often preferred  in semiconductor optics
for the perturbational treatment of the external
field. A summary closes our presentation.

\section{The dimerized tight-binding Hamiltonian}
\label{Hamiltpei}

Free, non-interacting spin-1/2 electrons in a {\em single\/} band do not show
optical absorption at finite frequencies because excitations
with vanishing momentum necessarily have vanishing energy.
All the optical conductivity is thus concentrated in the Drude
weight at zero frequency.
Hence, interactions must be included to obtain 
absorption at finite frequencies.

In highly anisotropic, quasi one-dimensional systems the lattice is unstable
against a static (Peierls-)distortion at low temperatures since the
gain in electron kinetic energy due to the distortion overcomes
the lattice deformation energy.
We assume that an electron transfer can take place between
neighboring atoms only (tight-binding approximation).
Furthermore, the wave vector of the lattice distortion,~$q$, is
taken commensurate with the
electron filling, i.e., $q=2k_F=\pi/a$ where $a$~is
the lattice spacing of the undimerized chain.
This implies that we address the half-filled case where
the number of lattice sites~$L$ equals the number of electrons,
$N=N_{\uparrow}+N_{\downarrow}=L$, $n=N/L=1$.

Under these assumptions the electrons are described by
a dimerized tight-binding model,
\begin{equation}
\hat{T}(\delta)= - t \sum_{l=1,\sigma}^{L}
\left(1+ (-1)^l\delta\right)
\left(
\hat{c}_{l,\sigma}^+ \hat{c}_{l+1,\sigma}^{\phantom{+}} 
+ \hat{c}_{l+1,\sigma}^+ \hat{c}_{l,\sigma}^{\phantom{+}}
\right) \; .
\end{equation}
Here, $\hat{c}_{l,\sigma}^+$ ($\hat{c}_{l,\sigma}^{\phantom{+}}$) 
are the creation (annihilation) operators
for an electron of spin~$\sigma$ in a Wannier
orbital centered at site~$l$.
$0 \leq \delta \leq 1$ describes
the effect of the bond-length alternation on the electron transfer
amplitudes.
It implies~$t=(t_1+t_2)/2$, and $\delta=(t_2-t_1)/(t_1+t_2)\geq 0$
if one denotes the
hopping amplitudes for the long (short) bond by~$t_1$ ($t_2$).
Usually, one supposes an exponential dependence of the
electron transfer amplitudes
on the distance~$r$ between sites, $t(r)=t_0 \exp\left(-(r-a)/\lambda\right)$.
For $r_1=a (1+\eta/2)$, $r_2=a (1-\eta/2)$ this implies $t_0=\sqrt{t_1t_2}$
and $\lambda=\eta a/\ln(t_2/t_1)$ where $\eta$
parametrizes the bond length alternation.
Both~$\delta$ and $\eta$ are usually
small, e.g., $\delta=0.2$ and $\eta=-0.06$ are standard values for
polyacetylene (Baeriswyl, Campbell, and Mazumdar 1992); 
(Schott and Nechtschein 1994). %cite
Note that $\delta$ and $\eta$ are always opposite in sign.

As usual the Hamiltonian can be diagonalized in momentum space.
We apply periodic boundary conditions, and introduce the
Fourier transformed electron operators
as $\hat{c}_{k,\sigma}^+=\sqrt{1/L} \sum_{l=1}^L
\exp(ikla) \hat{c}_{l,\sigma}^+$ for the~$L$ momenta~$k=2\pi m/(La)$,
$m=-(L/2),\ldots (L/2)-1$.
We may thus write
\begin{equation}
\hat{T}(\delta) = \sum_{|k|\leq \pi/(2a),\sigma}
\epsilon(k)\left(
\hat{c}_{k,\sigma}^+\hat{c}_{k,\sigma}^{\phantom{+}}
-
\hat{c}_{k+\pi/a,\sigma}^+\hat{c}_{k+\pi/a,\sigma}^{\phantom{+}}
\right)
-i \Delta(k)
\left( \hat{c}_{k+\pi/a,\sigma}^+\hat{c}_{k,\sigma}^{\phantom{+}}
-
\hat{c}_{k,\sigma}^+\hat{c}_{k+\pi/a,\sigma}^{\phantom{+}}\right)
\label{Tink}
\end{equation}
with the dispersion relation~$\epsilon(k)$ and hybridization
function~$\Delta(k)$ defined as
\begin{mathletters}
\begin{eqnarray}
\epsilon(k) &=& -2t\cos(ka) \\[3pt]
\Delta(k) &=& 2t\delta\sin(ka)\; .
\end{eqnarray}
\end{mathletters}%

The Hamiltonian can easily be diagonalized in $k$-space by introducing
the new Fermion quasi-particle operators $\hat{a}_{k,\sigma,\pm}^{+}$
for these two bands which are related to the original electron operators by
\begin{mathletters}
\label{mixamp}
\begin{eqnarray}
\hat{a}_{k,\sigma,-}^{\phantom{+}} &=&
\alpha_k \hat{c}_{k,\sigma}^{\phantom{+}} + i \beta_k
\hat{c}_{k+\pi/a,\sigma}^{\phantom{+}} \\[6pt]
\hat{a}_{k,\sigma,+}^{\phantom{+}} &=&
\beta_k \hat{c}_{k,\sigma}^{\phantom{+}} -i \alpha_k
\hat{c}_{k+\pi/a,\sigma}^{\phantom{+}}
\end{eqnarray}
\end{mathletters}%
for $|k|\leq \pi/(2a)$.
The inverse transformation reads
\begin{mathletters}
\label{mixamp2}
\begin{eqnarray}
\hat{c}_{k,\sigma}^{\phantom{+}} &=&
\alpha_k \hat{a}_{k,\sigma,-}^{\phantom{+}} + \beta_k
\hat{a}_{k,\sigma,+}^{\phantom{+}} \\[6pt]
\hat{c}_{k+\pi/a,\sigma}^{\phantom{+}} &=&
-i \beta_k \hat{a}_{k,\sigma,-}^{\phantom{+}} + i \alpha_k
\hat{a}_{k,\sigma,+}^{\phantom{+}} \; ,
\end{eqnarray}
\end{mathletters}%
where $\alpha_k^2+\beta_k^2=1$ has to be fulfilled for a canonical
transformation.
We note the helpful relations
\begin{mathletters}
\label{useful}
\begin{eqnarray}
% ONE
\hat{c}_{k,\sigma}^+\hat{c}_{k,\sigma}^{\phantom{+}}
-
\hat{c}_{k+\pi/a,\sigma}^+\hat{c}_{k+\pi/a,\sigma}^{\phantom{+}}
&=&
\left(\alpha_k^2-\beta_k^2\right)
\left(\hat{a}_{k,\sigma,-}^+\hat{a}_{k,\sigma,-}^{\phantom{+}}
-
\hat{a}_{k,\sigma,+}^+\hat{a}_{k,\sigma,+}^{\phantom{+}}\right)
\nonumber \\[3pt]
&& +
2\alpha_k\beta_k
\left(\hat{a}_{k,\sigma,-}^+\hat{a}_{k,\sigma,+}^{\phantom{+}}
+
\hat{a}_{k,\sigma,+}^+\hat{a}_{k,\sigma,-}^{\phantom{+}}\right)
\\[6pt]
%
%
% TWO
\hat{c}_{k,\sigma}^+\hat{c}_{k,\sigma}^{\phantom{+}}
+
\hat{c}_{k+\pi/a,\sigma}^+\hat{c}_{k+\pi/a,\sigma}^{\phantom{+}}
&=&
\hat{a}_{k,\sigma,-}^+\hat{a}_{k,\sigma,-}^{\phantom{+}}
+
\hat{a}_{k,\sigma,+}^+\hat{a}_{k,\sigma,+}^{\phantom{+}} \\[6pt]
%
%
% THREE
\hat{c}_{k,\sigma}^+\hat{c}_{k+\pi/a,\sigma}^{\phantom{+}}
-
\hat{c}_{k+\pi/a,\sigma}^+\hat{c}_{k,\sigma}^{\phantom{+}}
&=&
i \left(\alpha_k^2-\beta_k^2\right)
\left(\hat{a}_{k,\sigma,+}^+\hat{a}_{k,\sigma,-}^{\phantom{+}}
+
\hat{a}_{k,\sigma,-}^+\hat{a}_{k,\sigma,+}^{\phantom{+}}\right)
\nonumber
\\[3pt]
&& +
i 2\alpha_k\beta_k
\left(\hat{a}_{k,\sigma,+}^+\hat{a}_{k,\sigma,+}^{\phantom{+}}
-
\hat{a}_{k,\sigma,-}^+\hat{a}_{k,\sigma,-}^{\phantom{+}}\right)
\\[6pt]
%
%
% FOUR
\hat{c}_{k,\sigma}^+\hat{c}_{k+\pi/a,\sigma}^{\phantom{+}}
+
\hat{c}_{k+\pi/a,\sigma}^+\hat{c}_{k,\sigma}^{\phantom{+}}
&=&
-i \left(\hat{a}_{k,\sigma,+}^+\hat{a}_{k,\sigma,-}^{\phantom{+}}
-
\hat{a}_{k,\sigma,-}^+\hat{a}_{k,\sigma,+}^{\phantom{+}}\right)
\; .
\end{eqnarray}
\end{mathletters}%
They allow to write the Hamiltonian in momentum space, eq.~(\ref{Tink}),
as
\begin{eqnarray}
\hat{T}(\delta) &=& \sum_{|k|\leq \pi/(2a),\sigma}
\left(\hat{a}_{k,\sigma,+}^+\hat{a}_{k,\sigma,+}^{\phantom{+}}
-
\hat{a}_{k,\sigma,-}^+\hat{a}_{k,\sigma,-}^{\phantom{+}}\right)
\left[ -\epsilon(k) (\alpha_k^2-\beta_k^2) - \Delta(k)2\alpha_k\beta_k\right]
\nonumber \\[6pt]
 &&
 \phantom{\sum_{|k|\leq \pi/(2a),\sigma}}
+ \left(\hat{a}_{k,\sigma,+}^+\hat{a}_{k,\sigma,-}^{\phantom{+}}
+ \hat{a}_{k,\sigma,-}^+\hat{a}_{k,\sigma,+}^{\phantom{+}}\right)
\left[ \epsilon(k) 2\alpha_k\beta_k - \Delta(k) (\alpha_k^2-\beta_k^2)\right] 
\; .
\end{eqnarray}
We demand that the mixing terms vanish. This gives the following relation
between $\alpha_k=\cos\phi_k$ and $\beta_k=\sin\phi_k$
\begin{equation}
\tan(2\phi_k) = \frac{\Delta(k)}{\epsilon(k)} \; .
\end{equation}
The mixing amplitudes become
\begin{mathletters}
\label{appalphabeta}
\begin{eqnarray}
2\alpha_k\beta_k &=& - \frac{\Delta(k)}{E(k)}\\[6pt]
\alpha_k^2-\beta_k^2 &=& - \frac{\epsilon(k)}{E(k)}\\[6pt]
\alpha_k &=& \sqrt{ \frac{1}{2} \left( 1 - \frac{\epsilon(k)}{E(k)}
\right) }\\[6pt]
\beta_k &=& - \sqrt{ \frac{1}{2} \left( 1 + \frac{\epsilon(k)}{E(k)}
\right)} {\rm sgn}\left(\Delta(k)\right)
\end{eqnarray}
\end{mathletters}%
where ${\rm sgn}(x)$ is the sign function.
Note that~$\alpha_{k+\pi/a}= -i\beta_k$, $\beta_{k+\pi/a}=i\alpha_k$.
This implies that~$\alpha_{k+q}$ is complex, in general.

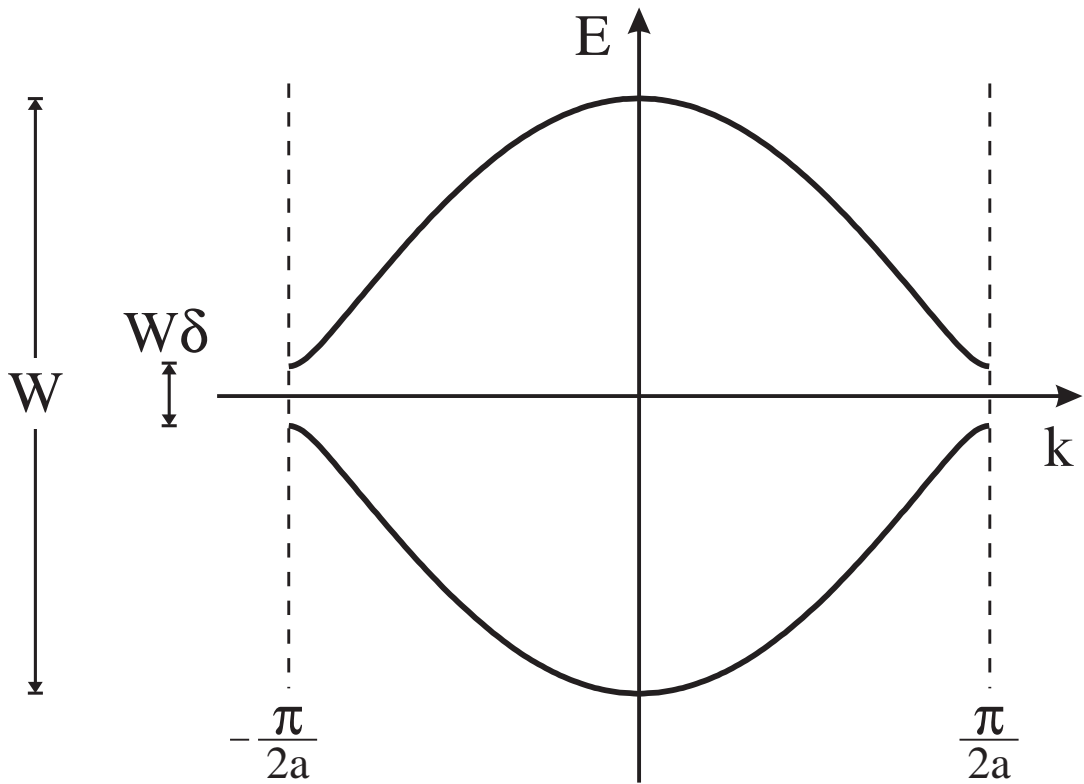
\begin{figure}[th]
%\vspace*{8cm}
\caption{Band structure of a Peierls insulator for $\delta=0.2$.
\hskip8cm}
\label{Hueckdis}
\end{figure}

It is easily seen that the Hamiltonian is diagonal
in the new operators. The result is
\begin{equation}
\hat{T}(\delta)
= \sum_{|k|\leq \pi/(2a),\sigma} E(k) (\hat{a}_{k,\sigma,+}^+
\hat{a}_{k,\sigma,+}^{\phantom{+}}
- \hat{a}_{k,\sigma,-}^+ \hat{a}_{k,\sigma,-}^{\phantom{+}})
\; .
\label{Tdiapeierls}
\end{equation}
Here, $\pm E(k)$ is the dispersion relation for the upper~($+$)
and lower~($-$) Peierls band,
\begin{equation}
E(k) = \sqrt{\epsilon(k)^2 + \Delta(k)^2} \; .
\label{peierlsen}
\end{equation}
The dispersion of both bands in the reduced zone scheme is shown
in figure~\ref{Hueckdis}. The band width is~$W=4t$.
The model describes a Peierls insulator at half-filling because all
states in the lower Peierls band are filled,
all states in the upper Peierls band
are empty, and there is the Peierls gap~$\Delta^{\rm P}=W\delta$
between the two bands.
Optical transitions at finite frequency are now possible
in the frequency range~$\Delta^{\rm P}
\leq \hbar \omega \leq W$. We set $\hbar=1$ from now on.

\section{Current operator and electrical conductivity}
\label{currentop}

To describe optical transitions we need to include the interaction with
an external electrical field which is only slowly varying
in space. Only a frequency dependent field will be considered
in the following.
In general, a one-dimensional system
shall be described by a time-independent
Hamilton operator.
Optical excitations are created
by the time-dependent perturbation (Maldague 1977); (Mahan 1990)   %cite
\begin{equation}
\hat{H}_{\rm int} = - \hat{\jmath} \cdot \frac{{\cal A}(t)}{c}
\label{Hintj}
\end{equation}
which couples the external
vector potential ${\cal A}(t)$
to the current operator~$\hat{\jmath}$ of the system ($c$~is the speed of
light).

It is not trivial to derive the current operator for a dimerized system.
We choose the Coulomb gauge and set the scalar potential to zero.
Then the electrical field~${\cal E}(t)$ is given by the vector
potential~${\cal A}(t)$
alone, ${\cal E}(t) = - (1/c) (\partial {\cal A}(t))/(\partial t)$, or, upon
Fourier transformation,
\begin{equation}
f(\omega) = \int_{-\infty}^{\infty} dt e^{i\omega t} f(t)
\quad ; \quad
f(t) = \int_{-\infty}^{\infty} \frac{d\omega}{2\pi} e^{-i\omega t} f(\omega)
 \; ,
\end{equation}
${\cal A}(\omega)/c={\cal E}(\omega)/(i\omega)$. Both fields are real
functions of time, so that they both fulfill $f(\omega)=f^*(-\omega)$,
$f={\cal A}, {\cal E}$.
The electrical field can be taken constant over lattice distances
since we are interested in optical excitations 
in the regime of electron volts. These energies are small compared
to atomic energies, and the corresponding wave vectors are small compared
to typical electron momenta.

The presence of the vector potential changes the amplitudes
for the electron transfer between neighboring unit cells.
In the tight-binding approximation they are given by
\begin{equation}
t_{l+1,l}({\cal A})= \int d{\rm\bf x} \Phi_{l+1}^*({\rm\bf x})
\left[ \frac{1}{2m} \left( {\rm\bf p} +\frac{e}{c}{\cal A} \right)^2
+V_{\rm atom}({\rm\bf x}) \right] \Phi_l({\rm\bf x})
\end{equation}
where $(-e)$ is the charge of an electron, and the term in square
brackets is the band structure Hamiltonian for a single electron.
$\Phi_l({\rm\bf x})$ are the Wannier functions centered at site~$l$.
We can apply a gauge transformation to formally eliminate the vector
potential from the band structure Hamiltonian. The Wannier functions now
acquire a field dependence,
\begin{equation}
\Phi_l({\rm\bf x}) \mapsto \widetilde{\Phi}_l({\rm\bf x}) =
\left[ \exp \left( i(e/c)
\int^{{\rm\bf R}_l} {\cal A}({\rm\bf x'}) d {\rm\bf x'}
\right)\right] \Phi_l({\rm\bf x})
\end{equation}
where ${\rm\bf R}_l$ is the space coordinate of site~$l$.
Since the vector potential is independent
of the position in the unit cell we may take the field dependence
outside the integral for the transfer matrix elements.
Now that the vector field is along the chain
we can write
\begin{equation}
t_{l+1,l}({\cal A})= \exp\left( -ie (R_{l+1}-R_{l})
{\cal A}/c \right) t_{l+1,l} \; .
\end{equation}
This is the well-known Peierls substitution.

Note that we tacitly {\em assumed\/} that the Wannier wave functions
remain undistorted albeit the atoms come closer to or farther away from
each other due to the distortion. This implies
$|R_{l+1}-R_l-a| \ll a$ as a requirement for the
validity of the tight-binding approximation.
Otherwise the atomic wave functions will be deformed which will ultimately
result in a bi-atomic, molecular structure.
In a dimerized system the distances between neighboring sites are
given by
\begin{equation}
R_{l+1}-R_{l}= a (1 +\eta (-1)^l ) \quad ; \quad R_l = a(l-(-1)^l\eta/2)
\end{equation}
where~$\eta$ is the relative distortion. 
Formally, we may allow $| \eta| \leq 1$ to check our results.
On physical grounds we have to require $|\eta| \ll 1$ to be consistent
with the tight-binding approximation.
The small value $\eta \approx -0.06$ in polyacetylen
(Baeriswyl, Campbell, and Mazumdar 1992); (Schott and Nechtschein 1994)  %cite
indicates that in real materials
the corrections due to the $\eta$-terms will dominate
those due to Wannier function deformations. The latter corrections
should be exponentially
small for small $|\eta|$-values since the wave functions themselves
decay exponentially as function of distance from the lattice site.

The Hamiltonian depends on the external field
via the field dependent hopping amplitudes.
We expand it in powers of
the (weak) external vector potential and obtain
\begin{equation}
\hat{T}(\delta,{\cal A}) = \hat{T}(\delta) - \hat{\jmath} \frac{{\cal A}}{c}
+ \frac{e^2a^2}{2} \hat{T}^{\prime}(\delta,\eta)
\left(\frac{{\cal A}}{c}\right)^2 + \ldots
\end{equation}
which allows us to identify the operator for the paramagnetic
{\em particle current\/},
$\hat{\jmath}$, and the operator for the diamagnetic {\em field current\/},
$\hat{T}^{\prime}(\delta,\eta)$, as
\begin{mathletters}
\begin{eqnarray}
\label{joperator}
\hat{\jmath} &=& (-e) \sum_{l,\sigma} (ita)
\left(
\hat{c}_{l+1,\sigma}^+ \hat{c}_{l,\sigma}^{\phantom{+}} 
- \hat{c}_{l,\sigma}^+ \hat{c}_{l+1,\sigma}^{\phantom{+}}
\right) \left(1+(-1)^l \delta\right) \left(1+(-1)^l \eta\right)
\\[6pt]
\hat{T}^{\prime}(\delta,\eta) &=&
(-t) \sum_{l,\sigma}  \left(
\hat{c}_{l+1,\sigma}^+ \hat{c}_{l,\sigma}^{\phantom{+}} 
+ \hat{c}_{l,\sigma}^+ \hat{c}_{l+1,\sigma}^{\phantom{+}}
\right) \left(1+(-1)^l \delta\right) \left(1+(-1)^l \eta\right)^2\; .
\end{eqnarray}
\end{mathletters}%
The operator for the time-dependent perturbation
theory in the external field is then identified as
$\hat{H}_{\rm int}(t) = -\hat{\jmath} \cdot {\cal A}(t)/c$
which is the standard form, equation~(\ref{Hintj}).
The geometrical distortion increases (decreases) the electron tunneling
between neighboring atoms ($\delta$-terms) but at the same time
decreases (increases) the size of the effective dipole ($\eta$-terms).

The operator for the total current, $\hat{\jmath}_{\rm tot}$,
is not only given by~$\hat{\jmath}$ because this quantity is not
gauge invariant. In fact, the particle current is supplemented by
the field current since the field itself carries momentum.
We write
\begin{equation}
\hat{\jmath}_{\rm tot} = \hat{\jmath} + \hat{X} {\cal A}/c
\label{one}
\end{equation}
and demand that this quantity be gauge invariant.
This fixes the unknown operator~$\hat{X}$.
Applying the same gauge transformation as above we can gauge
away the term proportional to~$\hat{X}$,
\begin{equation}
\hat{\jmath}_{\rm tot} \mapsto \hat{\jmath}({\cal A}) \; .
\label{two}
\end{equation}
We expand eq.~(\ref{one}) to first order in~${\cal A}$ and find from
eq.~(\ref{two}) that
$\hat{X}=e^2a^2\hat{T}^{\prime}(\delta,\eta)$, as expected.

\section{Optical conductivity}
\label{appeleccurr}

The dielectric function~$\widetilde{\epsilon}(\omega)$
and the coefficient for the linear optical
absorption~$\widetilde{\alpha}(\omega)$ are
given by (Haug and Koch 1990)   %cite
\begin{mathletters}
\begin{eqnarray}
\widetilde{\epsilon}(\omega) &=& 1 +\frac{4\pi i \sigma(\omega)}{\omega}
\label{epssigma}\\[6pt]
\widetilde{\alpha}(\omega) &=&
\frac{4\pi {\rm Re}\{\sigma(\omega)\}}{n_b c}
\end{eqnarray}
\end{mathletters}%
where ${\rm Re}\{\ldots\}$ denotes the real part and
$n_b$ is the background refractive index. 
It is supposed to be frequency independent near a resonance.
Hence, the real part of the optical conductivity
directly gives the absorption spectrum of the system.

Standard time-dependent perturbation theory gives the Kubo formula
for the optical conductivity (Maldague 1977); (Mahan 1990).    %cite
The Fourier transform of the total current {\em density\/}
and the Fourier transform of the electrical field~${\cal E}(\omega)$
define the optical conductivity
\begin{equation}
\sigma(\omega) = \frac{{\cal N}_{\perp}}{La}
\frac{\langle \hat{\jmath}_{\rm tot}(\omega)\rangle}{{\cal E}(\omega)}
= \frac{\chi(\omega) +
{\cal N}_{\perp} e^2 a \langle \hat{T}^{\prime}(\delta,\eta)\rangle/L}%
{i\omega}
\label{sigmachi}
\end{equation}
where $\langle \ldots \rangle$ means the expectation value
in the ground state without the perturbation, and ${\cal N}_{\perp}$
is the number of chains per unit area perpendicular to the chain direction.
Here, $\chi(\omega)$ is the current-current correlation
function defined by
\begin{equation}
\chi(\omega) = \frac{{\cal N}_{\perp}}{La}
i \int_0^{\infty} dt e^{i\omega t} \langle
\left[\hat{\jmath}(t),\hat{\jmath}\right]_- \rangle
\end{equation}
where $\hat{\jmath}(t)$ is the Heisenberg current operator 
for the unperturbed system.

The current-current correlation function can be spectrally
decomposed in terms of exact eigenstates~$| n\rangle$ (energy $E_n$)
of the unperturbed system as
\begin{equation}
\chi(\omega) = \frac{{\cal N}_{\perp}}{La}
\sum_n |\langle 0 | \hat{\jmath} | n\rangle|^2
\left[ \frac{1}{\omega +(E_n-E_0) +i\gamma} -
\frac{1}{\omega -(E_n-E_0) +i\gamma} \right] \; .
\label{decomp}
\end{equation}
Although $\gamma =0^+$ is infinitesimal we may introduce $\gamma>0$
as a phenomenological broadening
of the resonances at $\omega = \pm(E_n-E_0)$.
The real part of the optical conductivity determines the absorption
spectrum. It is given by
\begin{equation}
{\rm Re}\{ \sigma(\omega) \} = 
\frac{  {\rm Im}\{ \chi(\omega) \}   }{\omega}
= \frac{{\cal N}_{\perp} \pi}{La \omega}
\sum_n \left| \langle 0 | \hat{\jmath} | n\rangle \right|^2
\left[ \delta\left( \omega
-(E_n-E_0)\right) -\delta\left( \omega +(E_n-E_0)\right) \right]
\label{speccomp}
\end{equation}
which is positive for all~$\omega$.

\section{Optical absorption in Peierls insulators}
\label{peierlssec}

According to equation~(\ref{speccomp}) we have to
calculate the transition matrix 
elements~$\left|\langle n |\hat{\jmath}|0\rangle\right|^2$
between the ground state~$|0\rangle$
(energy~$E_0$) and all exact excited states~$|n\rangle$
(energy~$E_n$).
This is easily accomplished for non-interacting electrons.

The current operator in momentum space reads
\begin{eqnarray}
\hat{\jmath} &=& -e \sum_{l,\sigma}
ita \left(1+ (-1)^l\delta\right)\left(1+ (-1)^l\eta\right)
\left(
\hat{c}_{l+1,\sigma}^+ \hat{c}_{l,\sigma}^{\phantom{+}} 
- \hat{c}_{l,\sigma}^+ \hat{c}_{l+1,\sigma}^{\phantom{+}}
\right)
\nonumber 
\\[6pt]
&=& -e \left[ \sum_{|k|\leq \pi/a,\sigma}
(1+\eta\delta)\frac{\partial \epsilon(k)}{\partial k}
\hat{c}_{k,\sigma}^+\hat{c}_{k,\sigma}^{\phantom{+}}
- i \left(1+\frac{\eta}{\delta}\right)
\frac{\partial \Delta(k)}{\partial k}
\hat{c}_{k+\pi/a,\sigma}^+\hat{c}_{k,\sigma}^{\phantom{+}}
\right]  \; .
\label{currentb}
\label{current}
\label{currenta}
\end{eqnarray}
It is seen that the current cannot simply be derived from 
the $k$-derivative of the Hamiltonian~$\hat{T}(\delta)$ in
momentum space, eq.~(\ref{Tink}).
There are additional terms proportional to~$\eta$ due to the fact that
the distance between two lattice points is~$\Delta r=a(1\pm\eta/2)$.
This has been ignored in previous treatments
(Genkin and Mednis 1968); (Cojan, Agrawal, and Flytzanis 1977);
(Baeriswyl, Harbeke, Kiess, Meier, and Meyer 1983); (Abe 1993).    %cite

In terms of the new Fermions the
current operator can be split into two terms,
one which acts within the Peierls subbands while the other
induces transitions between them,
$\hat{\jmath}=
\hat{\jmath}_{\rm intra}^{\rm P} + \hat{\jmath}_{\rm inter}^{\rm P}$. We obtain
\begin{mathletters}
\label{jpeierls}
\begin{eqnarray}
\hat{\jmath}_{\rm inter}^{\rm P} &=& 
\sum_{|k|\leq \pi/(2a),\sigma} \lambda_{\rm inter}^{\rm P}(k)
\left(\hat{a}_{k,\sigma,+}^+\hat{a}_{k,\sigma,-}^{\phantom{+}}
+ \hat{a}_{k,\sigma,-}^+\hat{a}_{k,\sigma,+}^{\phantom{+}}\right)
\label{jpeierlsa}
\\[6pt]
\hat{\jmath}_{\rm intra}^{\rm P} &=& 
\sum_{|k| \leq \pi/(2a),\sigma} \lambda_{\rm intra}^{\rm P}(k)
\left(\hat{a}_{k,\sigma,-}^+\hat{a}_{k,\sigma,-}^{\phantom{+}}
- \hat{a}_{k,\sigma,+}^+\hat{a}_{k,\sigma,+}^{\phantom{+}}\right)
\label{jpeierlsb}
\end{eqnarray}
with the transition matrix elements
\begin{eqnarray}
\lambda_{\rm inter}^{\rm P}(k)
&=& ea \left[ \delta \frac{(2t)^2}{E(k)} + \eta E(k)\right]
\\[6pt]
\lambda_{\rm intra}^{\rm P}(k) &=& ea ( 1-\delta^2)
\frac{\Delta(k)\epsilon(k)}{\delta E(k)}
\; .
\end{eqnarray}
\end{mathletters}%
If we restrict ourselves to half-filling or, more generally, to
transitions for positive frequencies
(interband transitions)
we may ignore intraband transitions which only contribute to
the optical conductivity~$\sigma(\omega)$ at $\omega=0$.

Now it is very simple to calculate the optical absorption
with the help of the spectral decomposition
after the Hamiltonian has been diagonalized.
One obtains
\begin{mathletters}
\label{optabsPei}
\begin{eqnarray}
{\rm Re}\{\sigma(\omega >0)\}
&=& \frac{\pi {\cal N}_{\perp}}{La \omega} \sum_{|k|\leq  \pi/(2a),\sigma}
\left[\lambda_{\rm inter}(k)\right]^2 \delta(\omega - 2E(k))
\\[9pt]
&=& \frac{{\cal N}_{\perp} e^2a}{2\omega^2}
\frac{\left(\delta W^2+\eta\omega^2\right)^2}{\sqrt{\left(\omega^2-(W\delta)^2
\right)\left(W^2-\omega^2\right)\,}} \quad \hbox{\rm for}\quad
W\delta < \omega < W
\end{eqnarray}
\end{mathletters}%
where~$W=4t$ is the band width, and
${\cal N}_{\perp}$ is the number of chains per unit area
perpendicular to the chain direction.

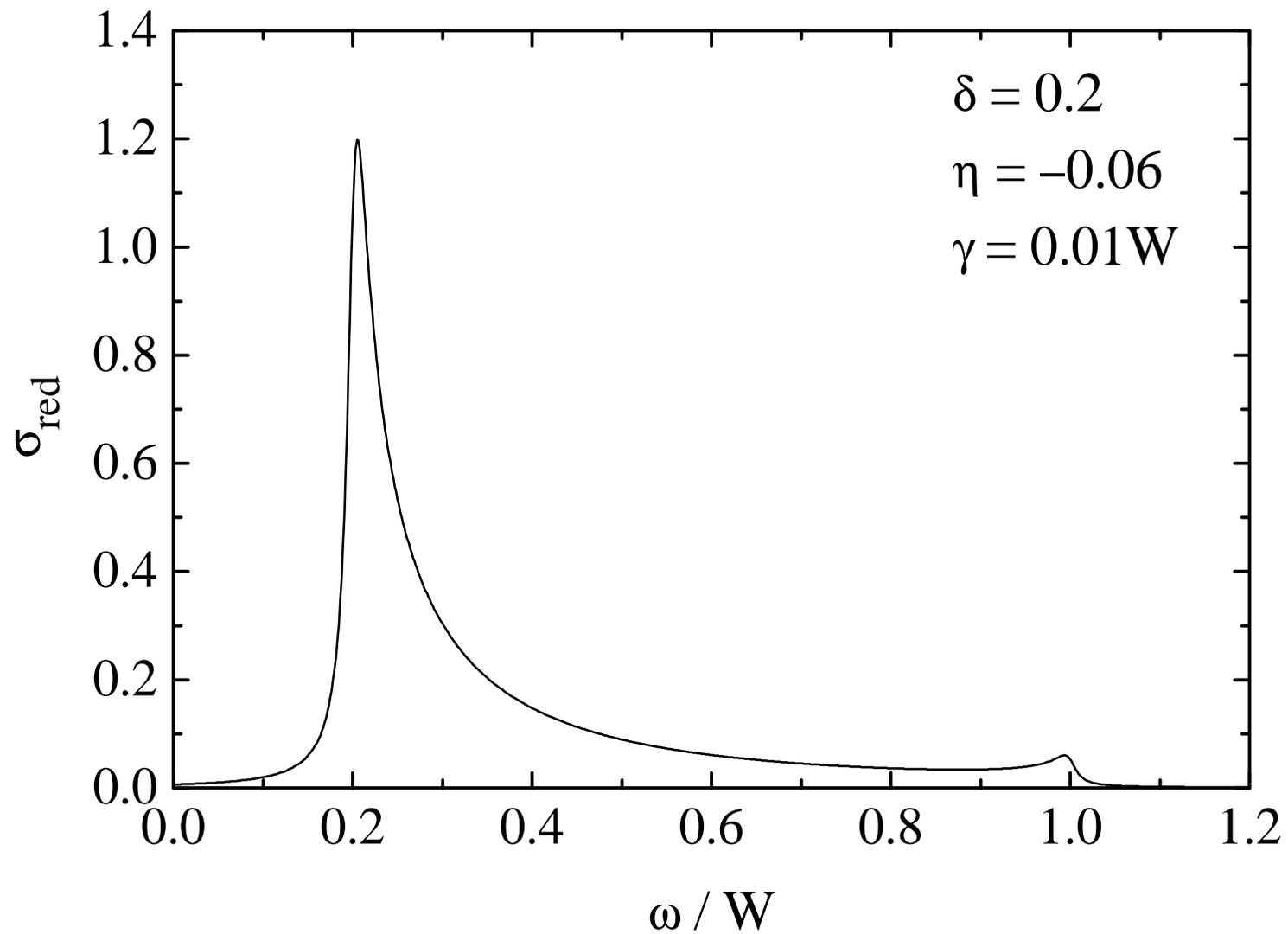
\begin{figure}[th]
%\vspace*{8cm}
\caption{Reduced optical conductivity, $\sigma_{\rm red}(\omega >0)$
in a Peierls insulator
for~$\delta=0.2$, $\eta=-0.06$.
A broadening of~$\gamma=0.01W$ has been included.}
\label{optabsPeifig}
\end{figure}

The optical absorption after eq.~(\ref{optabsPei}) is shown in
figure~\ref{optabsPeifig}.
We plot the dimensionless reduced optical conductivity
\begin{equation}
\sigma_{\rm red}(\omega >0) =
\frac{\omega {\rm Re}\{\sigma(\omega>0)\}  }%
{{\cal N}_{\perp}a e^2 W  } \; .
\label{sigmared}
\end{equation}
Furthermore we replace the energy conservation~$\delta(\omega-2E(k))$
by the smeared function
\begin{equation}
\widetilde{\delta}(x) = \frac{\gamma}{\pi(x^2+\gamma^2)} 
\end{equation}
to include effects of phonons and experimental
resolution. This is equivalent to the replacement $\omega \to \omega +i\gamma$
in the formulae for the spectral decomposition of
the optical conductivity.

The optical absorption~$\widetilde{\alpha}(\omega)$
is not only given by the joint density of states for
the upper and lower Peierls band but also reflects
the strong momentum dependence of the interband matrix element~$\lambda_{\rm
inter}(k)$.
The contribution of the van-Hove singularity at~$\omega=\delta W$ is thus
enhanced compared to the singularity at~$\omega=W$.
This is the well-known ``Umklapp enhancement'' (Philipps 1966).   %cite
The effect enforced by the contribution of the $\eta$-dependent terms
which have not been taken into account in previous works
(Genkin and Mednis 1968); (Cojan, Agrawal, and Flytzanis 1977);
(Baeriswyl, Harbeke, Kiess, Meier, and Meyer 1983); (Abe 1993).    %cite

\section{Electrical dipole operator}
\label{appdipole}

In semiconductor physics (Haug and Koch 1990) one often   %cite
prefers to calculate the
optical susceptibility~$\chi^{\rm opt}(\omega)$ which is related
to the dielectric function via
\begin{equation}
\widetilde{\epsilon}(\omega)= 1 + 4\pi \chi^{\rm opt}(\omega)
\; .
\end{equation}
Using eq.~(\ref{epssigma}) and~(\ref{sigmachi}) we find that
\begin{equation}
{\rm Im}\{\chi^{\rm opt}(\omega)\} = \frac{{\rm Im}\{\chi(\omega)\}}{\omega^2}
= \frac{{\cal N}_{\perp} \pi}{La}
\sum_n \frac{|\langle 0| \hat{\jmath}|n\rangle|^2}{\omega^2}
\left[ \delta\left( \omega
-(E_n-E_0)\right) -\delta\left( \omega +(E_n-E_0)\right) \right]
\; .
\label{justabove}
\end{equation}
We {\em define\/} the dipole operator~$\hat{P}$ via
\begin{equation}
\hat{\jmath}= - i \left[ \hat{P}, \hat{H}\right]_{-} 
\label{identify}
\end{equation}
and insert this definition into eq.~(\ref{justabove}).
This allows us to write
\begin{equation}
{\rm Im}\{\chi^{\rm opt}(\omega)\}
= \frac{{\cal N}_{\perp} \pi}{La}
 \sum_n |\langle 0| \hat{P}|n\rangle|^2
\left[ \delta\left( \omega
-(E_n-E_0)\right) -\delta\left( \omega +(E_n-E_0)\right) \right]
\end{equation}
which is the result for the optical absorption calculated
with the time-dependent perturbation operator
\begin{equation}
\hat{H}_{\rm int}(t) =  - \hat{P} {\cal E}(t)  \; .
\end{equation}
In principle we could have used this operator for the time-dependent
perturbation theory.

We can employ eq.~(\ref{identify}) to identify the dipole operator.
One finds
\begin{equation}
\hat{P} = -ea \sum_{l,\sigma}\left(l-(-1)^l\eta/2\right)
\hat{c}_{l,\sigma}^+\hat{c}_{l,\sigma}^{\phantom{+}}=
-e \sum_{l,\sigma} R_l \hat{c}_{l,\sigma}^+\hat{c}_{l,\sigma}^{\phantom{+}}
\end{equation}
which is indeed the result for the dipole operator in second quantization
for electrons at positions~$R_l$.
Unfortunately, this operator is sensitive to boundary
conditions and singular in the thermodynamical limit. Therefore, it cannot
be directly used in any practical calculation.

It is obvious that the dipole operator still contains a regular contribution
for {\em interband\/} transitions. To identify this
contribution we have to work in momentum space
and make use of eqs.~(\ref{Tdiapeierls}) and (\ref{jpeierlsa}). Then we
arrive at
\begin{mathletters}
\label{interPP}
\begin{eqnarray}
\hat{P}_{\rm inter}^{\rm P}&=& \sum_{|k|\leq \pi/(2a),\sigma}
\mu_{\rm inter}^{\rm P} (k)
\left(\hat{a}_{k,\sigma,+}^+\hat{a}_{k,\sigma,-}^{\phantom{+}}
- \hat{a}_{k,\sigma,-}^+\hat{a}_{k,\sigma,+}^{\phantom{+}}\right)
\\[6pt]
\mu_{\rm inter}^{\rm P} (k) &=&
-\frac{iea}{2}\left[ \delta \left(\frac{2t}{E(k)}\right)^2+\eta\right]
\end{eqnarray}
\end{mathletters}%
for the {\em interband\/} dipole operator 
(${\rm sgn}(\delta)=-{\rm sgn}(\eta)$).
Note that it is impossible to give the expression for an ``intraband
dipole operator'' because the $k$-th component
of the intraband current operator $\hat{\jmath}_{\rm intra}^{\rm P}$
of eq.~(\ref{jpeierlsb})
and the diagonalized Hamilton operator
$\hat{T}(\delta)$ of eq.~(\ref{Tdiapeierls})
are proportional to each other. In any case
such an ``intraband dipole operator'' would not give a contribution to
the optical (i.e., finite frequency) absorption, and can thus be ignored in
the case of a Peierls insulator.

In the form of equation~(\ref{interPP}) the interband dipole operator can
equivalently well be used for the calculation of the optical and higher
order susceptibilities for Peierls-distorted systems.

\section{Summary}

In this paper we have presented a detailed analysis
of the optical absorption in Peierls insulators. 
We observe the expected signature of a broad
band-to-band transition with van-Hove singularities at the absorption edges.
The strong momentum dependence of the transition matrix elements
suppresses the singularity at the high-frequency edge compared
to the one at the low-frequency edge. In contrast to previous
investigations we gave explicit analytical expressions for the optical
absorption and the dipole matrix elements for the tight-binding band structure.
In addition we found correction (``$\eta$''-)terms to the current operator
and dipole matrix elements which have been ignored in previous
works (Genkin and Mednis 1968); 
(Cojan, Agrawal, and Flytzanis 1977); 
(Baeriswyl, Harbeke, Kiess, Meier, and Meyer 1983); (Abe 1993).    %cite
They arise because the distance between two neighboring sites
is not constant but~$\Delta r=a(1\pm\eta/2)$.
However, they only slightly influence the final result for the optical
absorption. 

Unfortunately, a direct comparison to experiment is
difficult for two reasons. First, disorder effects can inhomogeneously
broaden single lines. Hence an experimentally
observed band can very well be a sign of disorder rather than
an argument for a Peierls insulator.
Secondly, when a residual electron-electron interaction
is included in a Peierls insulator 
(Abe, Yu, and Su 1992); (Abe, Schreiber, Su and Yu 1992)   %cite
one can equally well obtain excitons which draw the oscillator
strength form the band transitions. Hence, single exciton
lines on the other hand are not a clear-cut indication against
a Peierls insulator either.

\section*{Acknowledgments}

We thank H.~B\"{a}\ss ler, A.~Horv\'{a}th,
M.~Lindberg, S.~Mazumdar, M.~Schott, and
G.~Weiser for useful discussions.
The project was supported in part by the
Sonderforschungsbereich~383 
``Unordnung in Festk\"{o}rpern
auf mesoskopischen Skalen'' of the Deutsche Forschungsgemeinschaft.

\section*{References}
\begin{itemize}
\item S.~Abe, in: {\sl Relaxation in Polymers},
ed.~by T.~Kobayashi, (World Scientific, Singapore, (1993)), p.~215. 
\item S.~Abe, M.~Schreiber, W.~P.~Su, and J.~Yu, 
Phys.~Rev.~B~{\bf 45}, 9432 (1992).
\item S.~Abe, J.~Yu, and W.~P.~Su,
Phys.~Rev.~B~{\bf 45}, 8264 (1992).
\item L.~Alc\'{a}cer, and A.~Brau and J.-P.~Farges, in: {\sl Organic 
Conductors}, ed.~by J.-P.~Farges, (Marcel Dekker, New York, (1994)).
\item D.~Baeriswyl, G.~Harbeke, H.~Kiess, E.~Meier, and W.~Meyer,
Physica~B~{\bf 117-118}, 617 (1983). 
\item D.~Baeriswyl, D.~K.~Campbell, and S.~Mazumdar
in {\sl Conjugated Conducting Polymers}, ed.~by H.~Kiess,
(Springer Series in Solid State Sciences~{\bf 102},
Springer, Berlin (1992)).
\item C.~Cojan, G.~P.~Agrawal, and C.~Flytzanis,
Phys.~Rev.~B~{\bf 15}, 909 (1977).
\item J.-P. Farges (ed.), {\em Organic Conductors}, (Marcel Dekker, New York,
(1994)).
\item A.~Fritsch and L.~Ducasse, J.~Physique~I~{\bf 1}, 855 (1991).
\item F.~Gebhard, K.~Bott, M.~Scheidler, P.~Thomas, and S.W.~Koch, 
forthcoming articles.
\item V.~M.~Genkin and P.~M.~Mednis, Zh.~Eksp.~Teor.~Fiz.~{\bf 54},
1137 (1968); (Sov.\ Phys.~JETP~{\bf 27}, 609 (1968)).
\item H.~Haug and S.~W.~Koch, {\sl Quantum Theory of the
Optical and Electronic Properties of Semiconductors},
(World Scientific, Singapore, (1990)).
\item A.~J.~Heeger, S.~Kivelson, J.~R.~Schrieffer,
and W.-P.~Su, Rev.\ Mod.\ Phys.~{\bf 60}, 781 (1988).
\item G.~D.~Mahan, {\sl Many-Particle Physics}, (2nd~edition, Plenum
Press, New York (1990)).
\item P.~F.~Maldague, Phys.~Rev.~B~{\bf 16}, 2437 (1977).
\item S.~Mazumdar and S.~N.~Dixit, Phys.~Rev.~B~{\bf 34}, 3683 (1986).
\item F.~Mila, unpublished (1995).
\item J.~C.~Philipps, {\sl Solid State Physics}~{\bf 18},
ed. by F.~Seitz, D.~Turnbull, and H.~Ehrenreich, 55 (1966).
\item M.~Schott and M.~Nechtschein, in: {\sl Organic Conductors}, 
ed.~by J.-P.~Farges, (Marcel Dekker, New York, (1994)).
\end{itemize}

\typeout{figure captions}

\end{document}